# A Model for Randomly Correlated Deposition

**B. Karadjov and A. Proykova**

*Faculty of Physics, University of Sofia, 5 J. Bourchier Blvd. Sofia-1126, Bulgaria*
anap@phys.uni-sofia.bg

**Abstract**: A simple, discrete, parametric model is proposed to describe conditional (correlated) deposition of particles on a surface and formation of a connecting (percolating) cluster. The surface changes spontaneously its properties (phase transition) when a percolating cluster appears. The parameter **k** is included in the probability **p(k)** of particles to stick together and form a cluster on the surface. The case **k=1** corresponds to the ordinary 2D percolation on a square lattice. Thus the percolation threshold is controlled by the **k**-value: the larger **k** the higher threshold for percolation. The growth model seen from the percolation point of view allows us to describe several interesting applications in addition to irreversible aggregation in the presence of a repulsive force, **k>1**. For example, the occupied lattice sites might represent regions of specific magnetization in an otherwise disordered medium. Then the whole system is ordered or not according to the concentration of the deposited particles. Object-oriented code is developed for the Monte Carlo part of the calculations.

## I. Introduction

One way to study non-equilibrium interfaces is to construct discrete models and investigate them using computer simulations. The formation of a real surface is influenced by a large number of factors, and it is almost impossible to consider all of them. The hope is, however, that there is a small number of basic "laws" determining the morphology and the dynamics of the growth [1]. The action of these basic laws can be described in microscopic detail through discrete growth models - models that mimic essential physics bypassing some of the less essential details. Sometimes the interface growth is of interest because the surface must change its properties under given circumstances. For example, this is the critical phase transition. The critical change of conductance of covered surfaces could be considered in the frame of the percolation theory [2]. The conductivity is related to the percolation probability of the branching model, e.g. a network with all but a fraction of conducting links removed.

Percolation and phase transitions are well understood statistical phenomena [3]. The geometrical connectivity in the percolation problem [4] proved to be useful in understanding

The present paper presents a discrete, parametric model of *random correlated deposition* (RCD) of particles onto a clean surface, e.g. an empty square lattice **LxL**. The deposited particles stick conditionally together - a parameter **k** determines the conditional probability - and form aggregates (clusters), which might connect two opposite sides of a finite surface for a given particle concentration $\mathbf{p_c}$. The value of $\mathbf{p_c}$ is controlled by **k**. The most accurate value known for the unconditional percolation, **k=1**, is $\mathbf{p_c}$ = 0.59274621(13) obtained in Monte Carlo simulations of 7 billion samples of configurations obtained in a square lattice wrapped as a torus – no boundary conditions [6].

The goal of our project is to relate the percolation threshold $\mathbf{p_c(k)}$ to the repulsive force, e.g. **k**-value and to study the change of the fractal dimension $D_f$ of the connecting (percolating for $\mathbf{L}\rightarrow\infty$) cluster as a function of **k**. The classical percolating cluster is characterized with $D_f$ = **Error!** = 1.8958(3) [6].

## II. The model

Each simulation starts with an empty (clean) surface – a square lattice with **LxL** cells each labeled with *(i,j)*, *i,j* = 1,2,3,…L. A particle can stick (be deposited) to an empty, randomly selected cell depending on the state of the neighbouring cells. The rules for deposition are shown in Fig.1.

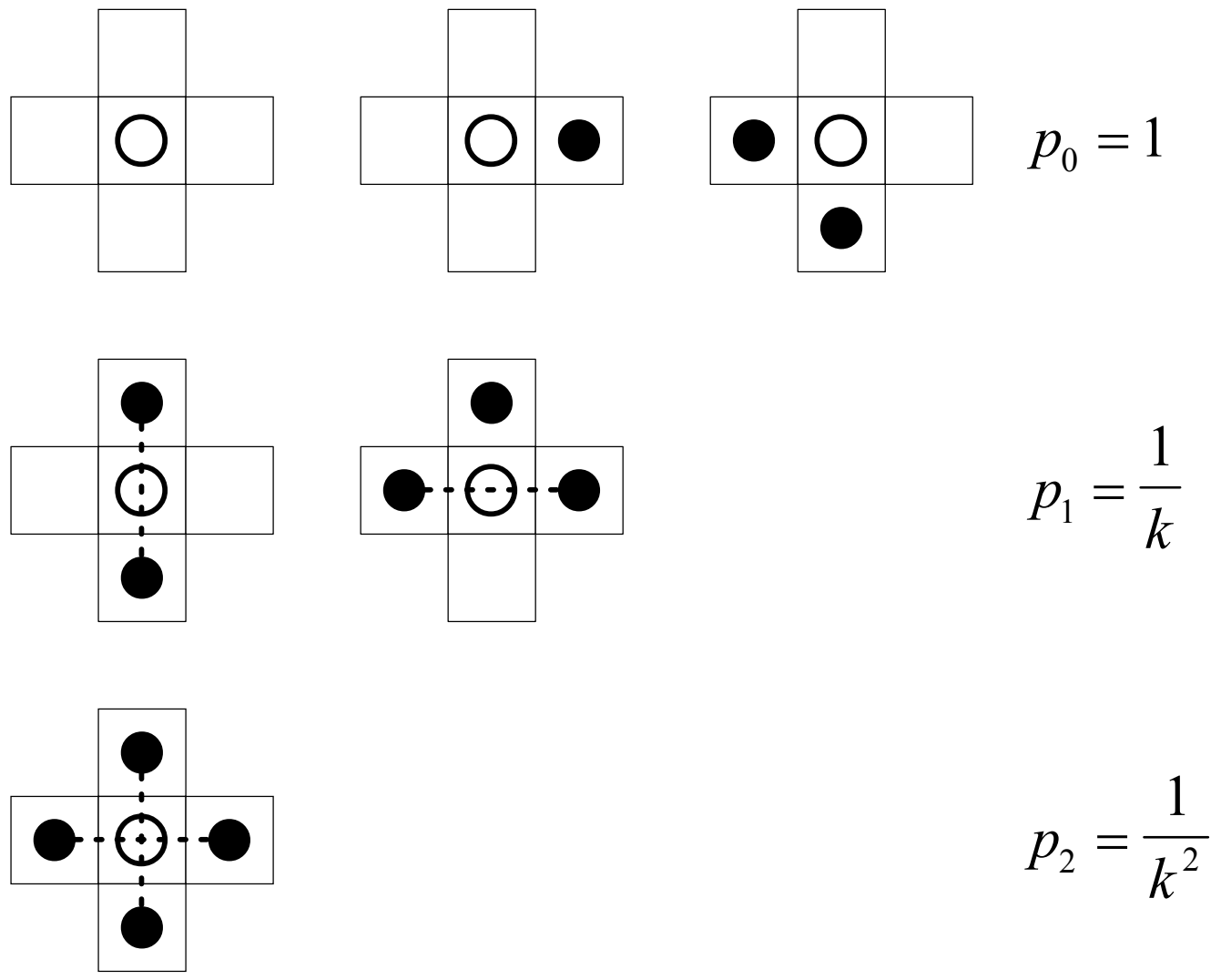

Fig.1 The model for correlated deposition of particles on a square surface.

The empty circle indicates the new particle which would be deposited with a probability 1 if there were no neighbours at the trial cell, one neighbour, or two neighbours along a diagonal. If three or two neighbours in up-down or left-right positions were there, then the probability was 1/k. The most repulsive are four neighbours and the probability for deposition is $1/k^2$. This model simulates a repulsive surface with a force depending on the local environment. Once a particle is deposited it is not allowed to diffuse. This is like the dust on the computer screen coated with a special anti-dust liquid.

It is obvious that **k=1** corresponds to the usual percolation on a square lattice – each particle is deposited independently on the status of the surrounding cells; repulsion is mimicked with **k > 1**, and the attraction with **k<1**.

## III. Results

We have studied the influence of **k** >1 on the percolation threshold, the fractal dimension and critical exponent for various sizes **L** = 50 – 1750. For a comparison we present the data for **k=1** as well. We have obtained 0.5928±0.0002 which is worse that the value reported in [6] probably due to the approach (Figure 2) we have followed for threshold determination. A finite-size lattice has a connecting rather than percolating (infinite) cluster. Formation of a connecting cluster (up-down and/or/xor left-right) changes the surface properties: clean becomes dirty, the isolator becomes a conductor. At low concentrations the probability of formation of a connecting cluster is small – it is zero for a number of particles smaller than the lattice side **L**. Following [7] we accept that the mean cluster size distribution is $P_p(p) = \Phi\left(\frac{p - p_c}{\sigma}\right)$, $\Phi(x) = \frac{1}{\sqrt{2\pi}}\int_{-\infty}^{x} e^{-\frac{y^2}{2}}\,dy$. Figure 2 illustrates the idea of determination the threshold $\mathbf{p_c}$ for a small lattice size, **L=250**. The dispersion σ limits the accuracy of the $\mathbf{p_c}$ –determination.

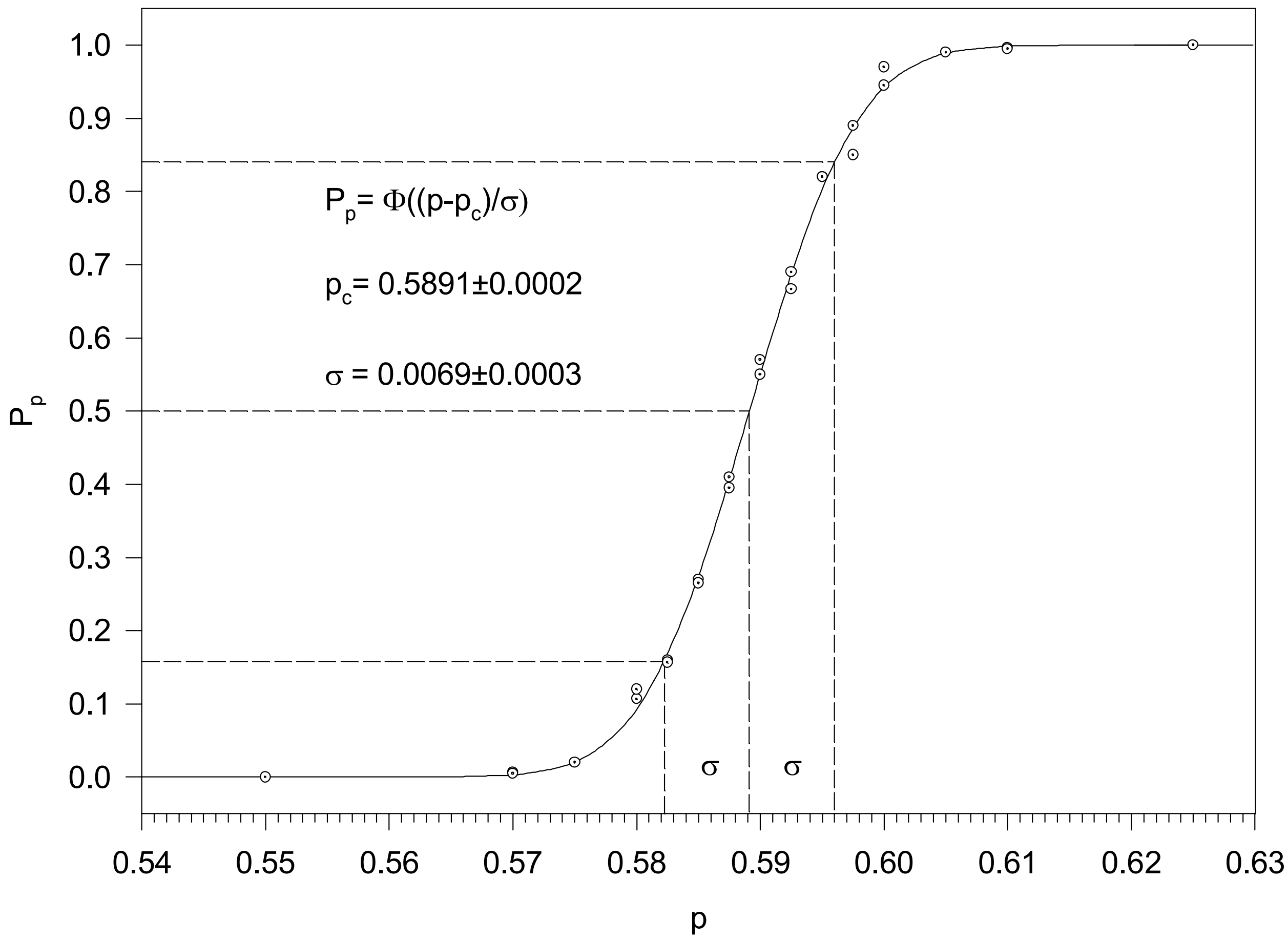

Figure 2 Determination of the percolation threshold from the simulations on a 250x250 square lattice, **k=1**.

The convergence of the threshold found for finites size lattices towards its actual value for $L \to \infty$ has been studied for various k. Recently a promising approach for finite-size corrections has been published [8].

Figure 3 a,b,c shows the size dependence of the percolation threshold computed from $p_c(L) = p_c(\infty) - \frac{C}{L^{\frac{1}{\nu}}}$ with an uncertainty $\sigma(L) \propto \frac{1}{L^{\frac{1}{\nu}}}$ [9].

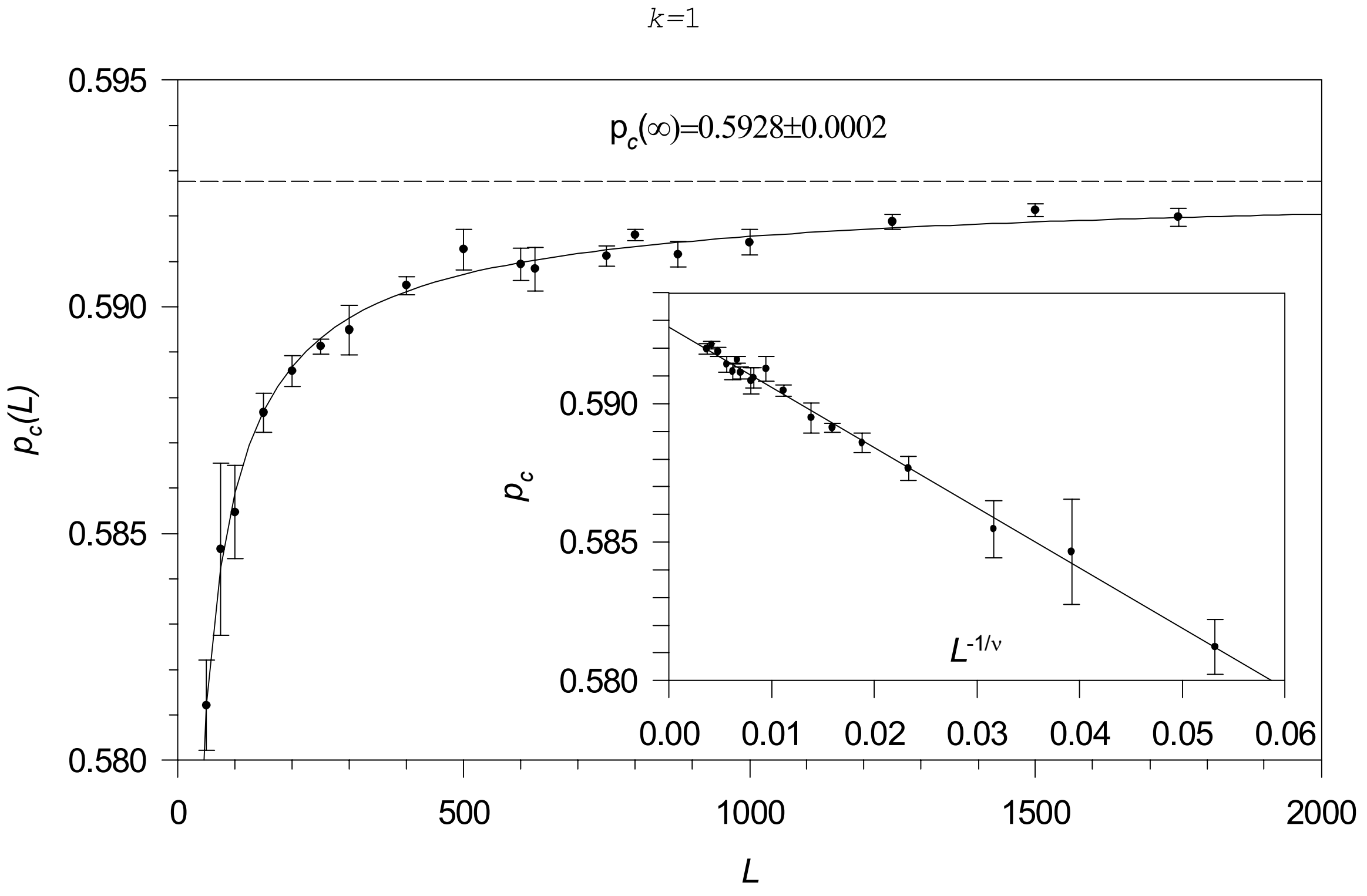

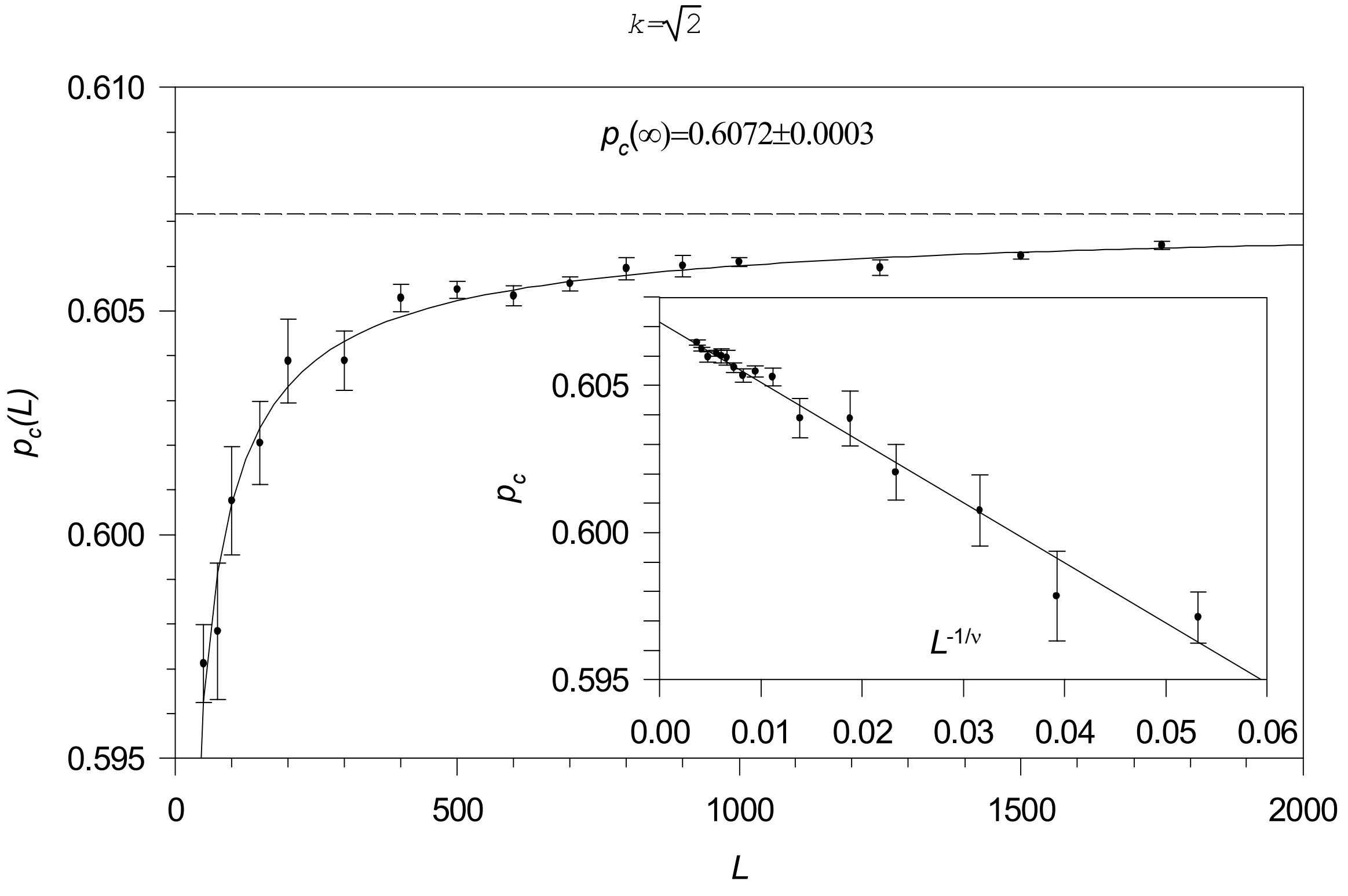

Fig.3a Determination of the percolation threshold for the case of **k=1** (the upper part of the figure, this is the usual percolation) and **k=√2**. The repulsion causes a shift towards higher threshold values.

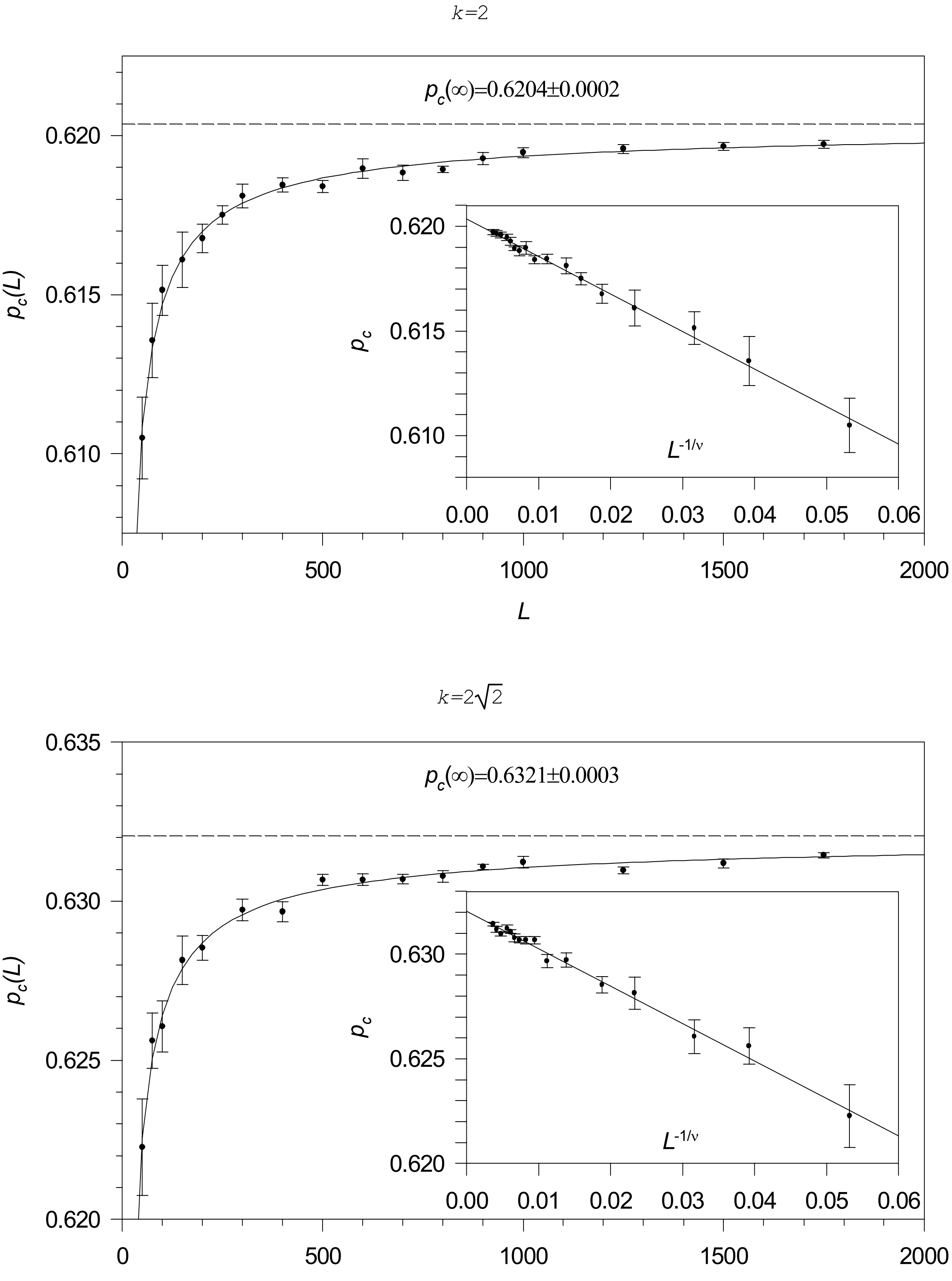

Fig. 3b Determination of the percolation threshold for the case of **k=2** (the upper part of the figure) and **k=2√2**. The repulsion causes a shift towards higher threshold values.

*k*=3

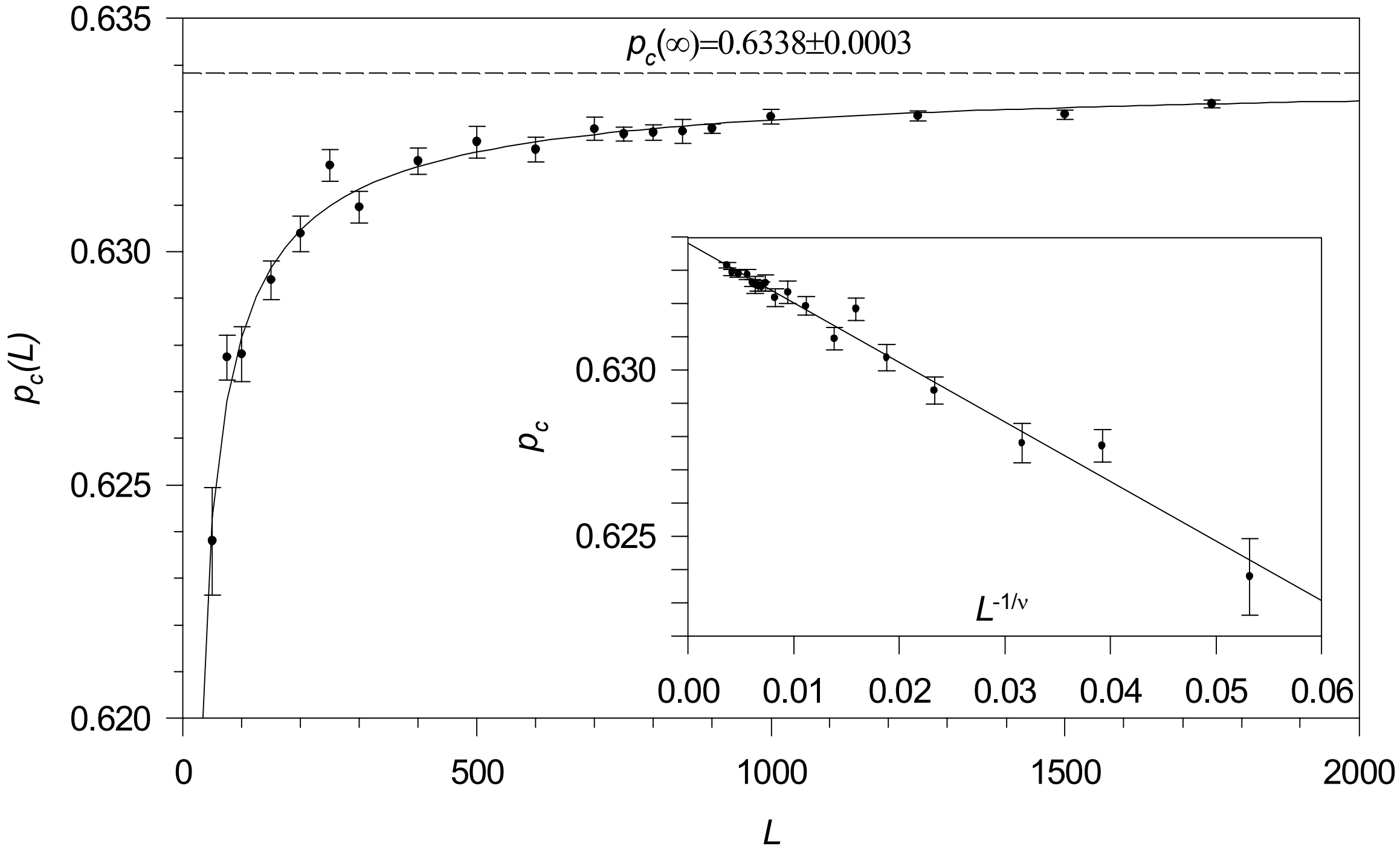

*k*=4

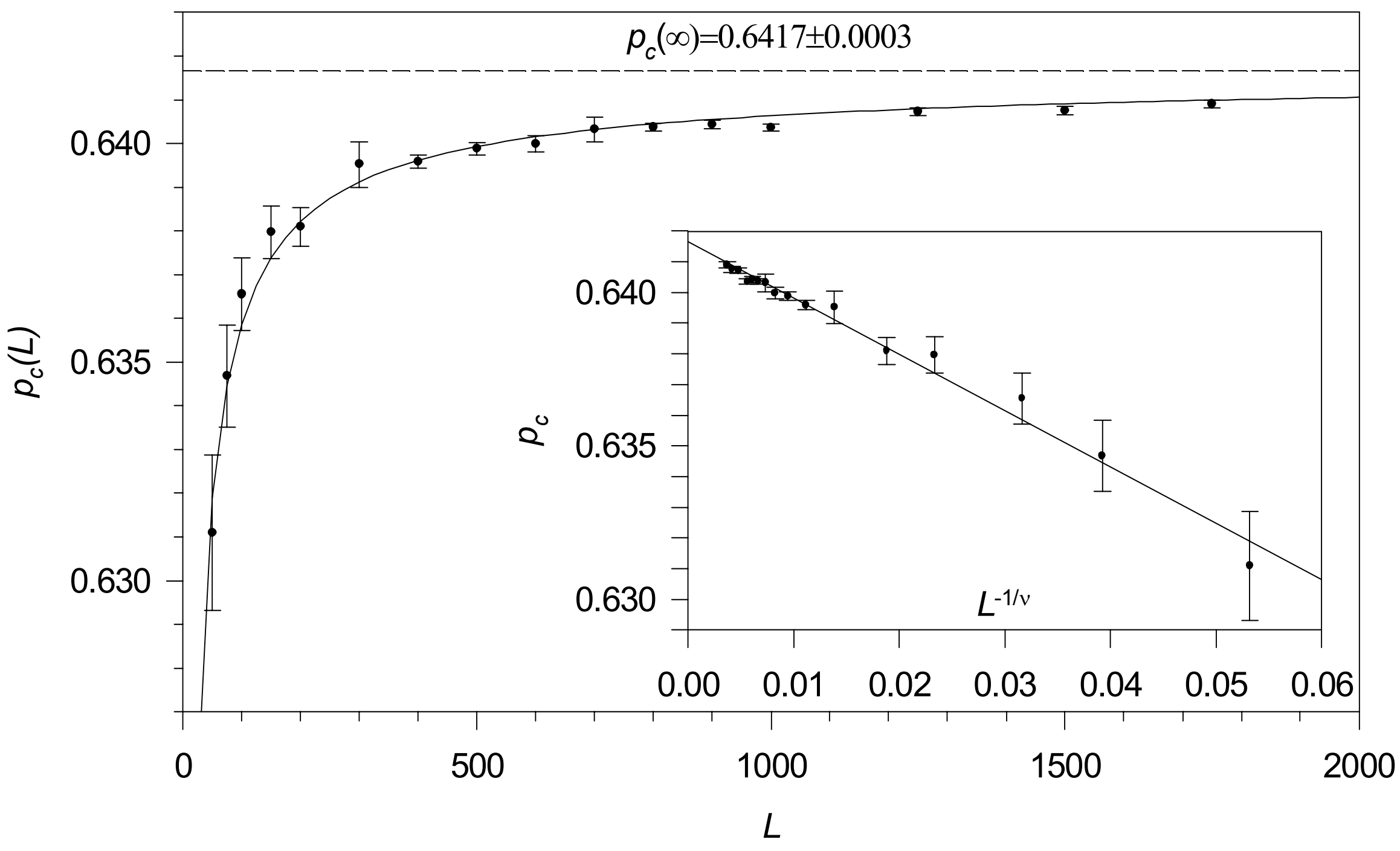

Fig. 3c Determination of the percolation threshold for the case of **k=3** (the upper part of the figure) and **k=4**. The repulsion causes a shift towards higher threshold values.

The table summarizes the results for the percolation threshold $\mathbf{p_c}$, its dispersion, the critical exponent, and the fractal dimension as functions of the repulsive force, **k**. The fractal dimension has been computed using a modification [10] of the box-counting technique [11].

| $K$ | $p_c$ | $\sigma_{100}$ | $\nu$ | $D_f$ |
|---|---|---|---|---|
| 1 | 0.5928±0.0002 | 0.0146±0.0006 | 1.41±0.07 | 1.895±0.002 |
| $\sqrt{2}$ | 0.6072±0.0003 | 0.0137±0.0004 | 1.33±0.06 | 1.898±0.002 |
| 2 | 0.6204±0.0002 | 0.0126±0.0004 | 1.36±0.05 | 1.892±0.001 |
| $2\sqrt{2}$ | 0.6321±0.0003 | 0.0120±0.0003 | 1.36±0.05 | 1.892±0.002 |
| 3 | 0.6338±0.0003 | 0.0114±0.0005 | 1.39±0.07 | 1.897±0.002 |
| 4 | 0.6417±0.0003 | 0.0109±0.0003 | 1.32±0.05 | 1.896±0.002 |

Let us compute the relative change of the threshold and its dispersion: $\Delta$ indicates the absolute change of the quantity for successive **k**-values; $\delta$ is the uncertainty. We get for the percolation threshold $\frac{\delta p_c}{\Delta p_c} \sim 10^{-2}$, and for the dispersion $\frac{\delta\sigma_{100}}{\Delta\sigma_{100}} \sim 1$, i.e. the effect is comparable with its error. However, a detailed analysis of the simulation data points at a decrease of the width in the vicinity of the percolation.

## IV. Conclusions

The presence of a 'repulsive' **k>1** force causes increase of the percolation threshold. However the critical exponent agrees with the well-known value (1.33) for 2D percolation and the fractal dimension is in accordance with the value for the ordinary percolation. This means that the only role of the repulsive force is to delay the percolation if we consider the process as a function of time for a fixed concentration value.

**Acknowledgment** This work has been financed by a special grant (F-3/2003) of the Ministry of Education and Science, Bulgaria, to support research at the universities.

**References:**

[1] Manna S. S. and Jan N., 1991, J. Phys. A 24 1593-1601.
[2] Stinchcombe, R., 1983, "Dilute Magnetism", in Domb, C., and Lebowitz, J. L., *Phase Transitions and Critical Phenomena,* Volume 7, Academic Press, 151-280.
[3] Stauffer D and Aharony A 1991 *Introduction to Percolation Theory* 2nd edn. (London: Taylor and Francis).
[4] Stanley H E and Ostrowsky N (eds.) 1990 *Correlations and Connectivity* (Dordrecht: Kluwer).
[5] M.E.J. Newman and R.M. Ziff, Phys. Rev. E64, 016706 (2001); Phys. Rev. Lett. 85, 4104 (2000).
[6] B. M. Smirnov, The properties of fractal clusters, Phys. Rep. **188**, 1 (1990).
[7] A.L. Efros, Physics and Geometry of Disorder (in Russian http://edu.ioffe.ru/edu/efros), eq.6.
[8] P.M.C. de Oliveira1, R.A. N´obrega1, and D. Stauffer, Brazilian Journal of Physics, vol. 33, no. 3, September, 2003.
[9] D. Stauffer, *Scaling theory of percolation clusters*, Physics Reports 54 1 (1979).
[10] B. Karadjov, Master thesis, University of Sofia (1996); in preparation for publication.
[11] B.B. Mandelbrot, *The Fractal Geometry of Nature*. New York, Freeman, 1983.